\documentclass[aps,prl,twocolumn,nofootinbib,preprintnumbers,showkeys]{revtex4-1}

\usepackage{graphicx}
\usepackage{dcolumn}
\usepackage{bm}
\usepackage{amsmath}
\usepackage{float}
\usepackage{multirow}
\usepackage[colorlinks=true, pdfstartview=FitV, bookmarks=true, bookmarksnumbered=true, breaklinks]{hyperref}

\usepackage{color}
\definecolor{blue}{rgb}{0.0, 0.0, 1.0}
\definecolor{red}{rgb}{1.0, 0.0, 0.0}
\definecolor{royalblue}{rgb}{0.0, 0.14, 0.4}
\hypersetup{linkcolor=red, citecolor=blue, urlcolor=royalblue}

\begin{document}

\title{Roper-like resonances with various flavor contents and their two-pion emission decays }
\author{A. J. Arifi$^{1}$}
\author{H. Nagahiro$^{1,2}$}
\author{A. Hosaka$^{1,3}$}
\author{K. Tanida$^{3}$}

\affiliation{
$^1$Research Center for Nuclear Physics (RCNP), Osaka University, Ibaraki, Osaka 567-0047, Japan\\
$^2$Department of Physics, Nara Women's University, Nara 630-8506, Japan\\
$^3$Advanced Science Research Center, Japan Atomic Energy Agency, Tokai, Ibaraki 319-1195, Japan}
\date{\today}

\begin{abstract}
We study the three-body decay of the newly observed bottom baryon $\Lambda_b^*(6072)$ by LHCb; 
$\Lambda_b^*(6072) \to \Lambda_b \pi \pi$. 
Its mass about 500 MeV above the ground state $\Lambda_b$ and a broad width imply that 
the state could be an analogue of the Roper resonance of the nucleon $N(1440)$.  
In terms of sequential processes going through  $\Sigma_b$ and $\Sigma_b^*$, 
we find that the observed invariant mass distribution is reproduced assuming its spin and parity $J^P = 1/2^+$. 
We discuss that the ratio of the two sequential processes and angular correlation of two pions
are useful for the determination of spin and parity.
We suggest further studies for the Roper resonance analogue in various flavor contents,  
raising an interesting and important question in baryon spectroscopy.  
\end{abstract}
\keywords{Roper-like resonance, flavor-independent nature, two-pion emission decay}

\maketitle

In this letter, motivated by the recent observation of $\Lambda_b^*(6072)$ by CMS and LHCb collaborations~\cite{Sirunyan:2020gtz,Aaij:2020rkw}, with mass and width the $M = 6072.3 \ \text{MeV}$ and $\Gamma = 72 \ \text{MeV}$ measured by LHCb~\cite{Aaij:2020rkw}, 
we discuss it as a state analogous to the Roper resonance $N(1440)$ with spin and parity $J^P = 1/2^+$ in heavy flavor sectors.  
It is shown that the three-body decay of two pion emission is particularly useful to determine its unknown spin and parity.  

The Roper resonance is the first excited state of the nucleon of $J^P = 1/2^+$~\cite{Roper:1964zza}. 
It has been a mysterious state because its properties such as the mass and level ordering with 
 negative parity nucleons have not been easily explained by the conventional quark model.
Recently, detailed studies by a dynamical model for meson-baryon scatterings~\cite{Suzuki:2009nj,Kamano:2010ud} 
and by the measurement of the transition form factor in a wide range of momentum transfer $Q^2$~\cite{Aznauryan:2009mx} 
have convinced that it is a radial excitation of the nucleon~\cite{Burkert:2017djo}.  

\begin{figure}[h]
\centering
\includegraphics[scale=0.49]{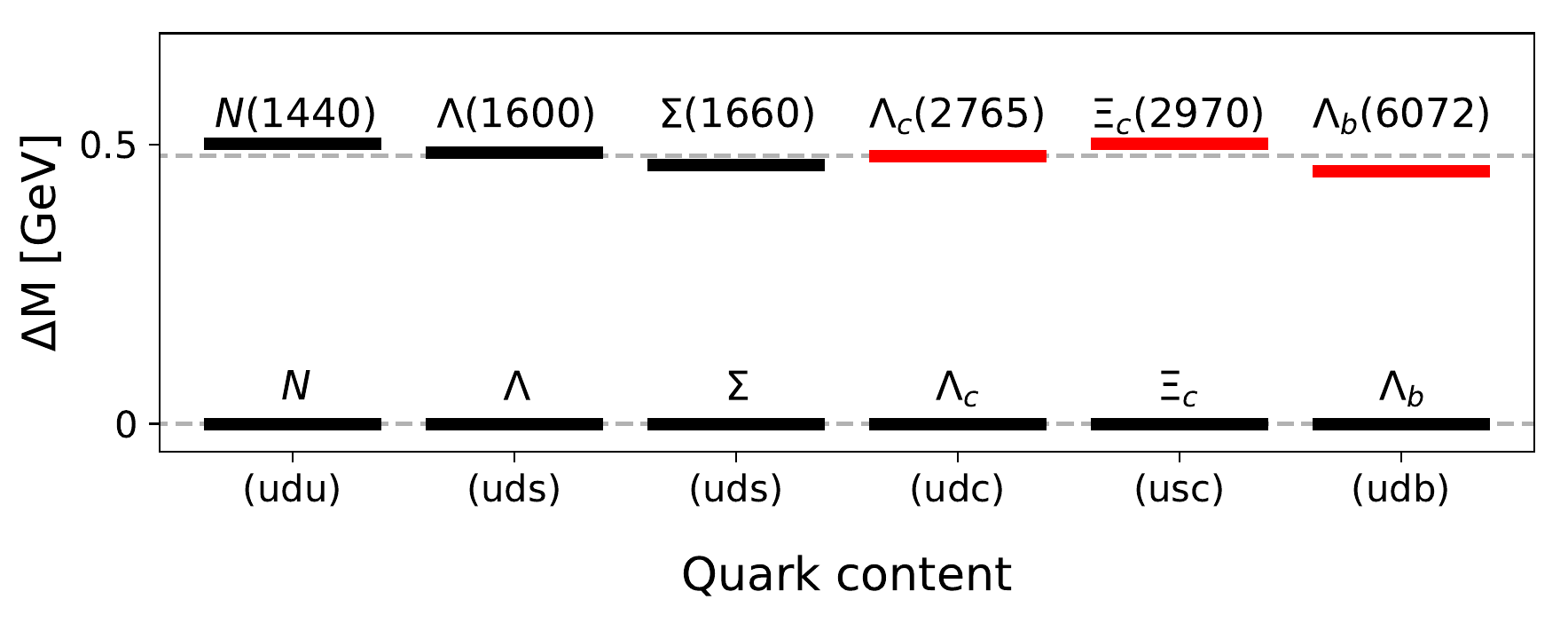}
\caption{\label{roper} Excitation energies of the first excited $J^P = 1/2^+$ baryons and candidates 
with various flavor contents. 
Red bars are for those with undetermined spin and parity.}
\end{figure}

On the other hand, similar states of $J^P = 1/2^+$ with almost the same excitation energies 
have been known for some time for hyperons systematically~\cite{Tanabashi:2018oca,Takayama:1999kc}.  
The recent observation of $\Lambda_b^*(6072)$ could add another candidate in the list of the analogous states.  
In fact, there is also a candidate in the charm sector, $\Lambda_c^*(2765)$, which was recently 
identified as an isoscalar particle~\cite{Abdesselam:2019bfp}.  
However, the spin and parity of these charmed and bottom candidates are not yet determined.  
Observing this situation, we show possible candidate states in Fig.~\ref{roper} with their
excitation energies.  
They are similar not only in masses but also in decays through two pion emission.  
These rather universal features in various flavor sectors
may suggest important dynamics of low energy QCD.  

\begin{figure}[h]
\centering
\includegraphics[scale=0.8]{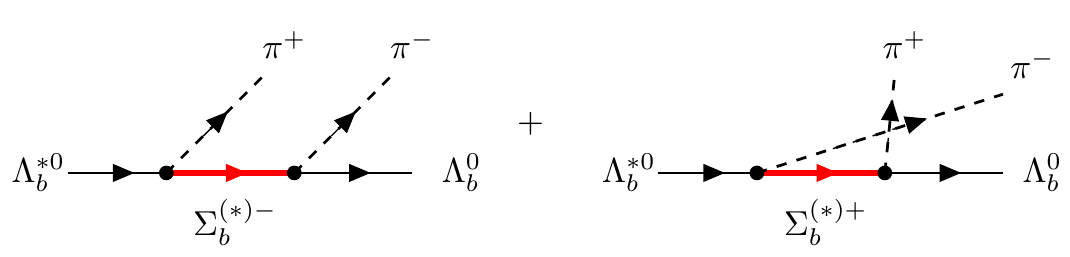}
\caption{\label{3body} Sequential decay processes of $\Lambda_b^{*0} \to \Lambda_b^0\pi^+\pi^-$ going through $\Sigma_b^{(*)-}$ and $\Sigma_b^{(*)+}$.}
\end{figure}

In the present letter, focusing on $\Lambda_b^*(6072)$, we study in detail  
its three-body decay $\Lambda_b^*(6072) \to \Lambda_b \pi \pi$.  
In our previous preprint~\cite{Arifi:2020ezz}, we have shown that the Dalitz plot analysis is particularly useful for the determination 
of spin and parity of $\Lambda_c^*(2765)$ when the decay is dominated by sequential processes going through 
$\Sigma_c$ and $\Sigma_c^*$. 
The relevant diagrams are shown in Fig.~\ref{3body} (for charmed baryons, 
replace the label $b$ for bottom by $c$).
Following this, we investigate 
the sequential decays of  $\Lambda_b^*(6072)$ with an intermediate state $\Sigma_b$ or $\Sigma_b^{*}$.
In principle, there is another one of the direct process where the two pions are emitted from a single vertex. 
As we will see later,  however, the sequential processes can explain the so far observed data very well.  

To compute these diagrams, let us first introduce amplitudes for various vertices. 
Because the bottom baryons are sufficiently heavy, we employ a non-relativistic method.  
Assuming that the spin and parity of $\Lambda_b^*$ equal $1/2^+$, the amplitudes at the first vertex of Fig.~\ref{3body}
are given by
\begin{eqnarray}
  -i\mathcal{T}_{\Lambda_b^* \to \Sigma_b \pi}   &=& g_1\ \chi^\dagger_{\Sigma_b} (\boldsymbol{\sigma} \cdot {\bf p}) \chi_{\Lambda_b^*}, \nonumber \\
  -i\mathcal{T}_{\Lambda_b^* \to \Sigma_b^* \pi} &=& g_2\ \chi^\dagger_{\Sigma_b^*} ({\bf S}^\dagger \cdot {\bf p}) \chi_{\Lambda_b^*},
  \label{eq_1stvertex}
\end{eqnarray}
and  for the second vertex, 
\begin{eqnarray}
  -i\mathcal{T}_{\Sigma_b \to \Lambda_b \pi} &=& g_3\ \chi^\dagger_{\Lambda_b} \left( \boldsymbol{\sigma} \cdot {\bf p}\right)  \chi_{\Sigma_b}, \nonumber \\
  -i\mathcal{T}_{\Sigma_b^* \to \Lambda_b \pi} &=& g_4\ \chi^\dagger_{\Lambda_b} \left( {\bf S} \cdot {\bf p}\right) \chi_{\Sigma_b^*}. 
\label{eq_2ndvertex}
\end{eqnarray}
In Eqs.~(\ref{eq_1stvertex}) and (\ref{eq_2ndvertex}), $g_i$'s are the coupling constants 
and $\chi$'s are the two-component spinors for the baryons as indicated by lower indices.

By using these vertex amplitudes, 
the amplitude of the first diagram of Fig.~\ref{3body} 
with the intermediate $\Sigma_b^-$ is expressed as
\begin{eqnarray}
-i\mathcal{T} \left[\Sigma_b^-\right] = - i \frac{\mathcal{T}_{\Sigma_b^- \rightarrow \Lambda_b^0\pi^-}\mathcal{T}_{\Lambda_b^{*0} \rightarrow \Sigma_b^-\pi^+} }{m_{23}-m_{\Sigma_b^-}+\frac{i}{2}\Gamma_{\Sigma_b^-}}, \quad \quad \label{amp1}
\end{eqnarray}
where $m_{23}$ is the invariant mass of the subsystem $\pi^-$(particle 2) and $\Lambda_b^0$(particle 3). 
The amplitudes with the $\Sigma_b^+$ intermediate state and those with $\Sigma_b^{* \mp}$
are computed similarly.  
The total amplitude is then the coherent sum of the four of them.  

The unknown coupling constants $g_i$'s are constrained 
by the heavy quark spin symmetry of QCD~\cite{Isgur:1991wq}; 
the ratios of $g_1$ to $g_2$ and of $g_3$ to $g_4$ are computed 
when $\Sigma_b$ and  $\Sigma_b^*$ are regarded as heavy quark spin partners.  
In the heavy quark limit, the spin $J$ of a baryon is composed of a spin $j$ of the brown muck of light degrees of freedom~\cite{Neubert:1993mb}
and spin 1/2 of the heavy quark, such that $J = j \pm 1/2$.  

For a spin 1/2 $\Lambda_b^*$, the brown muck spin can take two values,  $j = 0, 1$.  
Assume further that the brown muck of $\Sigma_b^{(*)}$ has spin $j=1$ as in the quark model, 
their total spin $J$ can be 1/2 or 3/2.  
Having this information, we can compute the ratio as
\begin{eqnarray}
\frac{g_2}{g_1}\ \  = \ \  \sqrt{2} \  {\rm for} \  j = 0 \ \ \ {\rm and } \ \ \ 
\frac{1}{\sqrt{2}} \ {\rm for} \ j = 1.
\end{eqnarray}
Combined with two-body phase space factors, 
when the same partial $L$-wave is possible for the two final states $\pi \Sigma_b$ and $\pi \Sigma_b^{*}$,
we find the ratio 
\begin{eqnarray}
R= \frac{\Gamma(\Lambda_b^*(6072)\to \Sigma_b^*\pi)}{\Gamma(\Lambda_b^*(6072)\to \Sigma_b \pi)}
=
 \frac{g_2^2}{g_1^2}  \frac{p_2^{2L+1}}{p_1^{2L+1}}.
 \label{ratios}
\end{eqnarray}

For $J^P = 1/2^+$, the two decays are in $p$-wave, $L = 1$.
Using $p_1 = 215$ MeV and $p_2 = 192$ MeV, we obtain
\begin{eqnarray}
R\ \  = \ \ 1.43  \ \ {\rm for} \  j = 0 \ \ \ {\rm and } \ \ \ 
0.36 \ {\rm for} \ j = 1
\end{eqnarray}
As we will see shortly, this agrees well with the observed data from LHCb.  
On the other hand, if $J^P = 1/2^-$ the partial waves of $\pi \Sigma_b$ and 
of $\pi \Sigma_b^{*}$ are $s$ and $d$-waves, respectively.  
For low momentum pion as in the present case, the $d$-wave decay is suppressed 
as compared to the $s$-wave decay.  
Therefore, we find  $R \ll 1$ which does not seem consistent with the experimental data.
This demonstrates the advantage in the use of the ratio $R$ for spin and parity.

\begin{figure}[b]
\centering
\includegraphics[scale=0.55]{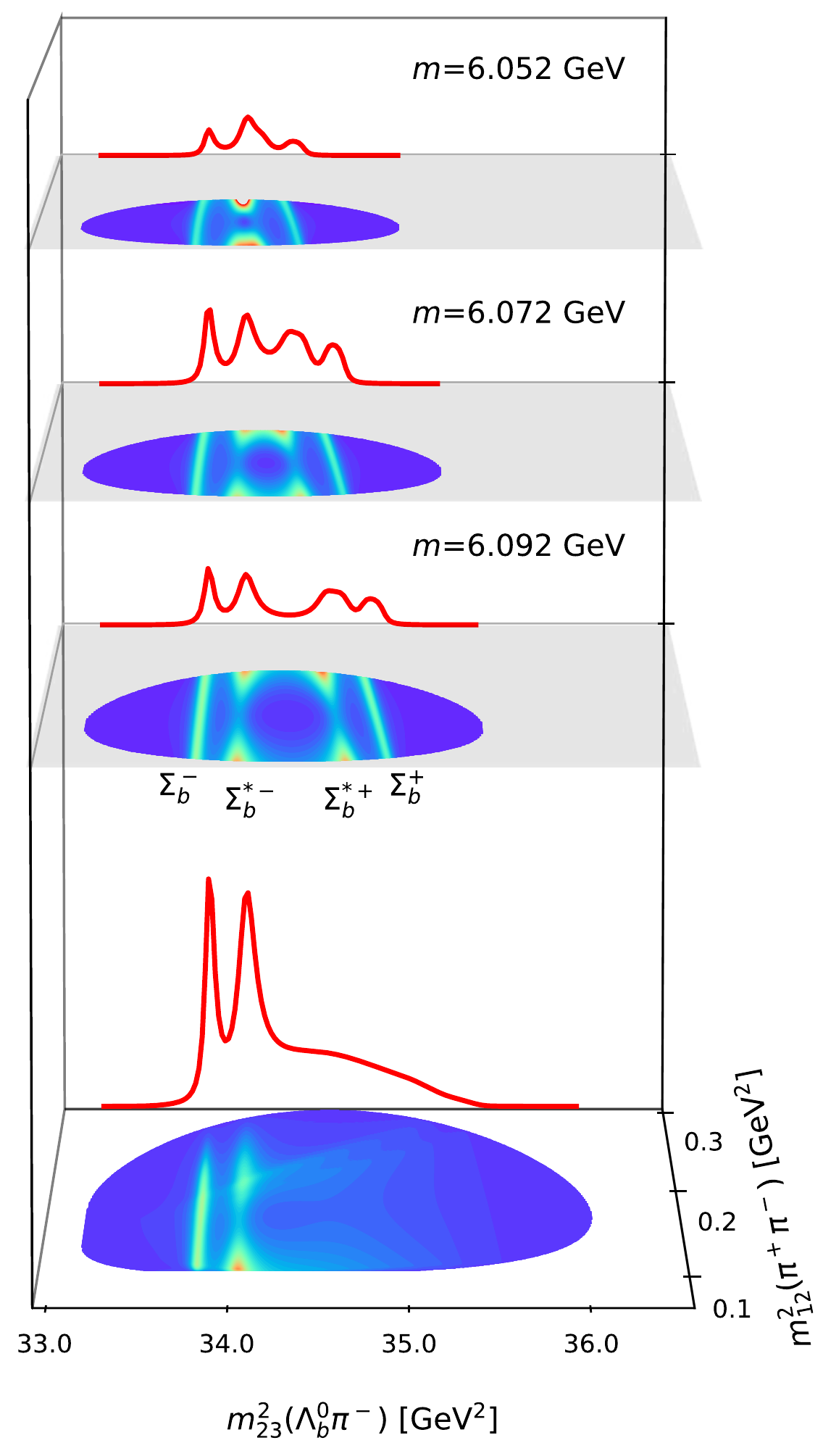}
\caption{\label{conv_dalitz} The Dalitz and invariant mass plots at three different initial masses of $\Lambda_b^{*}(6072)$ (upper three panels) and the convoluted one (bottom panel).}
\end{figure}

\begin{figure}[b]
\centering
\includegraphics[scale=0.55]{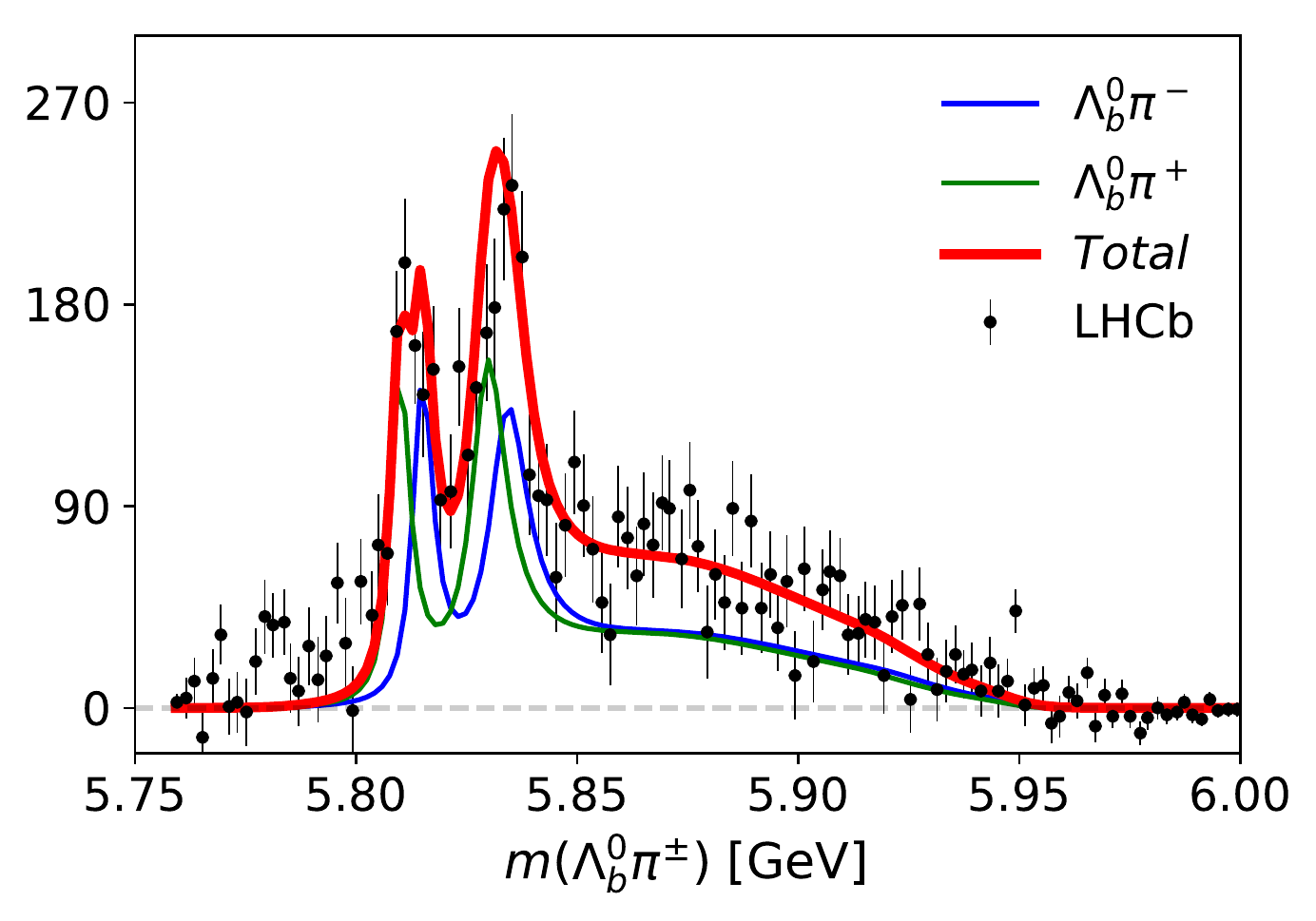}
\caption{\label{comparison} 
Invariant mass plots of $\Lambda_b^0\pi^-$ (blue), $\Lambda_b^0\pi^+$ (green) and their sum (red).  
The data from LHCb is the sum of these two~\cite{Aaij:2020rkw} and compared with the red line. 
}
\end{figure}

Now we discuss Dalitz plots and other related quantities.
In what follows, we assume again $J^P = 1/2^+$ for $\Lambda_b^*(6072)$.  
For decays of unpolarized particles, Dalitz plots are expressed as 
functions of two kinematical variables formed by the three 
decaying particles.  
In fact, they also depend on the mass of the initial particle, 
which distributes over a finite range of  
width $\sim$ 72 MeV for $\Lambda_b^*(6072)$.  
In experiments with low statistics, plots are often made by integrating the signals 
from such a finite mass range.   
Therefore, to compare directly theory predictions with the actual data, this feature is properly treated.  

In the upper three panels of Fig.~\ref{conv_dalitz} shown are diagonal views of the Dalitz plots for decay probabilities
at three different fixed masses of $\Lambda_b^*(6072)$ around the peak value, 
$m({\Lambda_b\pi\pi)} = $ 6.052, 6.072 and 6.092 GeV, 
as functions of invariant mass squares 
$m_{12}^2$ and $m_{23}^2$, 
$P(m; m_{12}^2, m_{23}^2)$.  
Note that the invariant mass $m_{23}$ is for $\Lambda_b \pi^-$
that couples to $\Sigma_b^{(*)-}$, which will be relevant in the discussions below.  
In each layer, the solid red line shown in the back panel is the projected strength, namely the invariant mass plot.  
In all the three plots, we find four resonance bands corresponding to $\Sigma_b^-$, $\Sigma_b^{*-}$, 
$\Sigma_b^{*+}$ and $\Sigma_b^{+}$ from the left to right.  
The first two are for the first diagram of Fig.~\ref{3body} showing the genuine resonance band, while the latter two 
correspond to the second one that are the so called kinematical reflection.  

Now what we need is the sum of the plots like these three ones.  
More precisely, the convolution is performed over the finite mass range weighted by the Breit-Wigner function, 
\begin{eqnarray}
P(m_{12}^2, m_{23}^2)  = \frac{1}{N} \int  \frac{P(m; m_{12}^2, m_{23}^2) \ {\rm d}m}{(m- M_{\Lambda_b^*} )^2 + \Gamma_{\Lambda_b^*}^2/4},
\end{eqnarray}
where the factor $N$ is for normalization and $m$ for the $\Lambda_b\pi\pi$ mass.  
The result is shown in the bottom panel of Fig.~\ref{conv_dalitz} as well as the invariant mass plot on the back panel.  
As shown there, in the convoluted plot the peaks of kinematical reflections in the upper three panels are smeared out, 
while the real resonance peaks remain.  

Because the experimental data is shown for the sum of signals of $\Lambda_b \pi^+$ and $\Lambda_b \pi^-$, 
we have shown in Fig.~\ref{comparison} the corresponding one
of our calculation (red solid line) compared with the experimental data. 
Note that experimental resolution of around 1 MeV is not considered in the calculation.
For theory side, we have also shown the strengths of $\Lambda_b \pi^-$ and $\Lambda_b \pi^+$ separately 
by blue and green solid lines, respectively.  
We find that our theory calculation agrees remarkably well
with experimental data not only in overall shape of the peaks and background.
The relative strengths of the peaks is determined by the ratio $R$, while the background shape 
is reproduced by the convoluted kinematical reflection.  
It is interesting to see that even such a detailed structure, the relative height of the two resonance peaks of
$\Sigma_b$ and $\Sigma_b^*$ are well reproduced.
This is achieved by taking the sum of the two charged states.
If only  $\Lambda_b \pi^-$  is included, the left peak is slightly higher than that of the right one 
as shown in the bottom panel of Fig.~\ref{conv_dalitz} and by the green line in Fig.~\ref{comparison}.  
This analysis so far supports the spin and parity of $\Lambda_b^*(6072)$ is $1/2^+$, and their decay 
is dominated by the two pion emission of the sequential processes.  
This is one of the main conclusions of the present study.  
If there will be further high statistics data, it is interesting 
to compare the Dalitz plots at each different energies.  

\begin{figure}[t]
\centering
\includegraphics[scale=0.46]{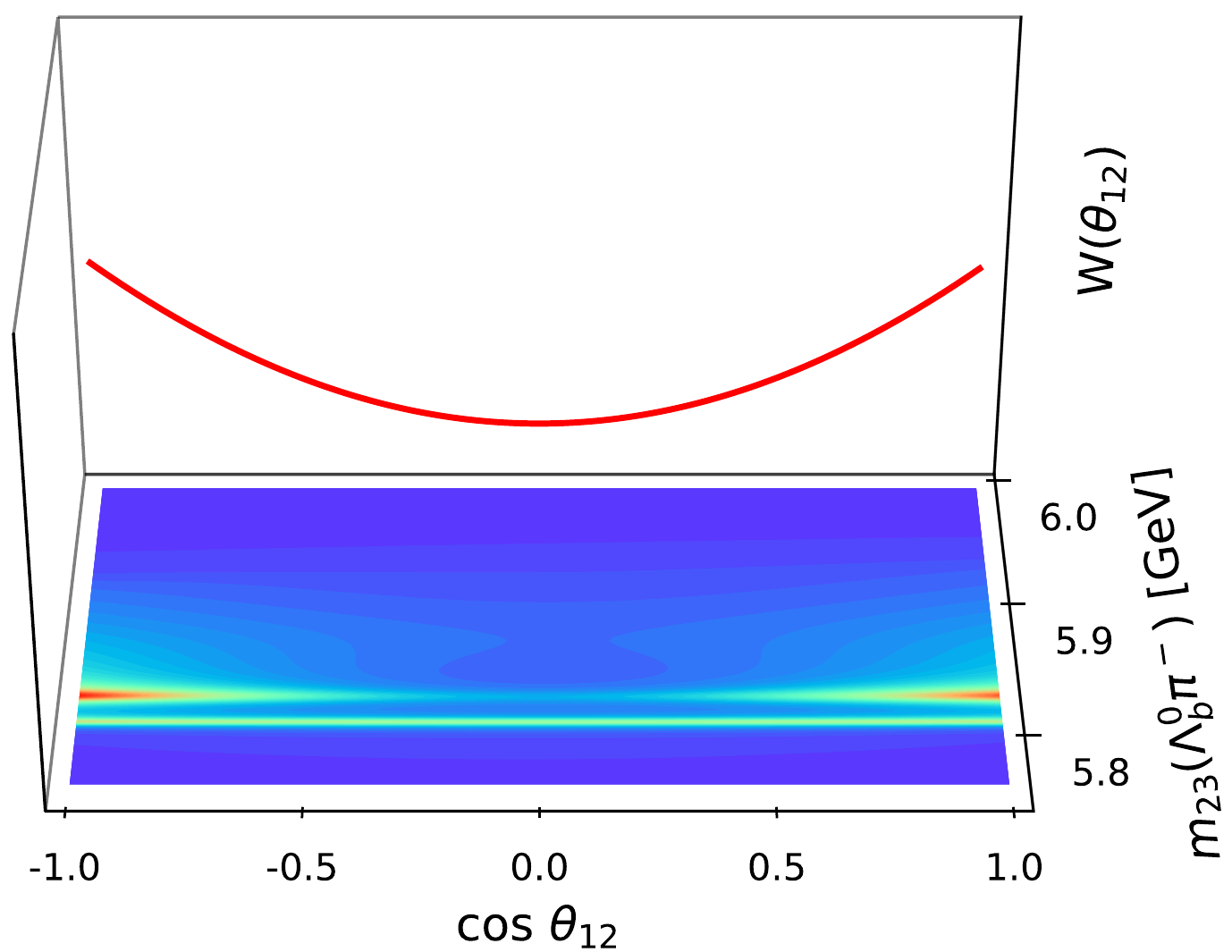}
\caption{\label{correlation} The convoluted square Dalitz plot for $\Lambda_b^*(6072)$ and its corresponding angular correlation along $\Sigma_b^{*-}$ with a mass cut $M_{\Sigma_b^{*-}} \pm \Gamma_{\Sigma_b^{*-}}$.}
\end{figure}

Let us further look at the angular correlation (dependence) 
along the resonance band.  
The angle is the one between the two pions $\theta_{12}$ in the intermediate resonance rest frame. 
To see the correlation better, it is convenient to make a square plot as a function 
of $m_{23}$ and $\cos \theta_{12}$ as shown in the lower plot of Fig.~\ref{correlation}.  
As seen there the angular correlation along the $\Sigma_b$ band (near side) is flat due to its spin 1/2.  
On the other hand, the angular correlation along the $\Sigma_b^*$ band (far side) has a concave 
shape due to its spin 3/2. 

In general the angular correlation along the $\Sigma_b^*$ band decaying from a particle of spin $J$ 
contains the two terms weighted by the helicity amplitudes $A_h$, where $h$ is the conserved helicity in the decay, 
\begin{eqnarray}
W(\theta_{12}) \propto && \left|A_{1/2}(\Lambda_b^* \to \Sigma_b^*\pi)\right|^2 \times (1 + 3 \cos^2\theta_{12}) \nonumber\\
		&+& \left|A_{3/2}(\Lambda_b^* \to \Sigma_b^*\pi)\right|^2 \times 3\sin^2\theta_{12}. \quad \quad \label{sin}
\end{eqnarray}
When the initial $\Lambda_b^*$ has spin 1/2, the helicity amplitude $A_{3/2}$ vanishes.
Thus, the angular correlation will have a $1 + 3 \cos^2\theta_{12}$ dependence (concave shape).
In reality, interference between the resonance of a finite width and background (kinematical reflection in the present case) 
contaminates that angular correlation.  
Fitting the projected angular correlation as plotted in the upper panel of 
Fig.~\ref{conv_dalitz}, our prediction for the angular correlation is 
\begin{eqnarray}
W(\theta_{12}) &\propto& 1 + 3.3\cos^2\theta_{12}.  
\end{eqnarray}

So far we have focused our discussions on $\Lambda_b^*(6072)$.
As anticipated in the beginning of this paper, there are other candidate baryons of similar nature, 
in particular $\Xi$ baryons.
One is $\Xi_c^*(2970)$ with excitation energy around 500 MeV and decays into $\Xi_c\pi\pi$~\cite{Yelton:2016fqw}.
If it is regarded as a Roper-like resonance, its decay into $\Lambda_c K$ is forbidden due to 
the brown muck selection rule, namely the transition of $0^+$ (initial brown muck) $\to 0^+$ (final brown muck) $ + 0^-$ (kaon)
does not occur.  
This selection rule applies to the other candidates and is a unique feature of the Roper like resonance.  

In recent analysis, LHCb reported several resonances in $\Lambda_c K$~\cite{Aaij:2020yyt}.
One of them is $\Xi_c^*(2965)$ with similar excitation energy, but it has significantly smaller decay width than  $\Xi_c^*(2970)$.
Therefore, $\Xi_c^*(2965)$ could be a particle of different nature that can be identified as a negative parity excitation 
of $\Xi_c'$ as pointed out in Ref.~\cite{Yang:2020zjl,Lu:2020ivo,Wang:2020gkn}.
Likewise, in the bottom sector,  the $\Xi_b^*(6227)$ resonance observed in $\Lambda_b K$ and $\Xi_b\pi$ decays~\cite{Aaij:2018yqz} 
would be this analogue of negative parity excitation of $\Xi_b^\prime$~\cite{Cui:2019dzj,Chen:2018orb,Wang:2018fjm,Yang:2019cvw,Jia:2019bkr}.
We expect a Roper like $\Xi_b^*$ of $J^P = 1/2^+$ decaying into  $\Xi_b\pi\pi$  at around 6.2 GeV.

We may expect further candidates in the strangeness sector.  
For example, the resonance $\Xi^*(1820)$ listed in PDG~\cite{Tanabashi:2018oca} may be identified with $3/2^-$ resonance as 
observed in $\Lambda K$ invariant mass~\cite{Biagi:1986vs}. 
However, a resonance with a similar excitation energy is also observed in $\Xi\pi\pi$ invariant masses 
with a larger decay width~\cite{Apsell:1970uf}.
This is a candidate of the Roper like resonance.  

Here we summarize several common features of the Roper-like resonances in our observation:
\begin{itemize}
\item The excitation energy is around 500 MeV which is rather flavor-independent.
\item The decay width is rather large with a significant coupling to two-pion emission decay.
\item Their two-pion emission decays are dominated by the sequential processes going through $1/2^+$ resonance, {\it e.g.} $\Sigma_b$ and $3/2^+$ resonance, {\it e.g.} $\Sigma_b^*$, or their analogues, indicating that the direct contributions including from $f_0(500)$ are insignificant as observed also for other baryons~\cite{Prakhov:2004zv,Prakhov:2004ri,Abe:2006rz}.
\item The ratio $R$ and angular correlation can be used to determine their spin and parity $1/2^+$.
\end{itemize}

Observations in the heavy flavor baryons and revisit to the light flavor baryons could provide new insights into 
unique flavor independent dynamics of low energy QCD.  

We thank support from the Reimei Research Promotion project (Japan Atomic Energy Agency) in completion of this work.
A. J. A. is supported by a scholarship from the Ministry of Education, Culture, Science and Technology of Japan.
H. N. is supported in part by Grants-in Aid for Scientific Research, Grants No. 17K05443(C) and A. H. by Grants No. 17K05441(C) and by Grants-in Aid for Scientific Research on Innovative Areas (No. 18H05407).

\nocite{*}
\bibliography{apssamp}

\end{document}